\RequirePackage{lineno}
\documentclass[aps,prd,onecolumn,twocolumn,showpacs,superscriptaddress,groupedaddress]{revtex4-2}
\usepackage{graphicx}  
\usepackage{dcolumn}   
\usepackage{bm}       
\usepackage{amssymb}   
\usepackage{booktabs}
\usepackage{longtable}
\usepackage{overpic}
\usepackage{multirow}
\usepackage{amsmath}
\usepackage{color}
\usepackage{epstopdf}
\usepackage[colorlinks,linkcolor=blue,anchorcolor=blue,citecolor=blue]{hyperref}
\usepackage{subfigure}
\usepackage{tabularx}
\usepackage{gensymb}
\usepackage{enumerate}
\usepackage{setspace}
\begin{document}
\begin{sloppypar}

\title{Cross section measurement of $e^{+}e^{-} \to f_{1}(1285)\pi^{+}\pi^{-}$ at center-of-mass energies between $3.808$ and $4.951\rm GeV$}
\author{
M.~Ablikim$^{1}$, M.~N.~Achasov$^{4,c}$, P.~Adlarson$^{76}$, O.~Afedulidis$^{3}$, X.~C.~Ai$^{81}$, R.~Aliberti$^{35}$, A.~Amoroso$^{75A,75C}$, Q.~An$^{72,58,a}$, Y.~Bai$^{57}$, O.~Bakina$^{36}$
% ...
\\
\vspace{0.2cm}
(BESIII Collaboration)\\
\vspace{0.2cm} {\it
$^{1}$ Institute of High Energy Physics, Beijing 100049, People's Republic of China\\
% ...
$^{3}$ Bochum  Ruhr-University, D-44780 Bochum, Germany\\
$^{4}$ Budker Institute of Nuclear Physics SB RAS (BINP), Novosibirsk 630090, Russia\\
% ...
$^{35}$ Johannes Gutenberg University of Mainz, Johann-Joachim-Becher-Weg 45, D-55099 Mainz, Germany\\
$^{36}$ Joint Institute for Nuclear Research, 141980 Dubna, Moscow region, Russia\\
% ...
$^{57}$ Southeast University, Nanjing 211100, People's Republic of China\\
$^{58}$ State Key Laboratory of Particle Detection and Electronics, Beijing 100049, Hefei 230026, People's Republic of China\\
% ...
$^{72}$ University of Science and Technology of China, Hefei 230026, People's Republic of China\\
% ...
$^{75}$ University of Turin and INFN, (A)University of Turin, I-10125, Turin, Italy; (B)University of Eastern Piedmont, I-15121, Alessandria, Italy; (C)INFN, I-10125, Turin, Italy\\
$^{76}$ Uppsala University, Box 516, SE-75120 Uppsala, Sweden\\
% ...
$^{81}$ Zhengzhou University, Zhengzhou 450001, People's Republic of China\\

\vspace{0.2cm}
$^{a}$ Deceased\\
% ...
$^{c}$ Also at the Novosibirsk State University, Novosibirsk, 630090, Russia\\
% ...
}

~\\
}

\date{\today}

\begin{abstract}
Using data samples collected by the \mbox{BESIII} detector located at the Beijing Electron Positron Collider, the cross sections of the process $e^+e^-\to f_{1}(1285)\pi^+\pi^-$ are measured at forty-five center-of-mass energies from $3.808$ to $4.951~{\rm GeV}$. An investigation on the cross section line shape is performed, and no significant structure is observed.
\end{abstract}
\maketitle

\section{\label{sec:level1}Introduction}
Over the past decades, a number of exotic candidates have been discovered in the charmonium and charmonium-like spectra. Excited vector charmonium states, which can be produced directly from $e^+e^-$ annihilation, have been extensively studied in both open-charm and hidden-charm final states. A notable example is the $Y(4260)$, initially observed by the BaBar experiment~\cite{BaBar:2005hhc} as a single broad peak in the $e^+e^-\to\gamma_{ISR} \pi^+\pi^- J/\psi$ process, where ISR denotes initial state radiation.~Later, the \mbox{BESIII} experiment performed a dedicated scan for the $e^+e^-\to \pi^+\pi^- J/\psi$ process, revealing that the state previously identified as the $Y(4260)$ consists instead of two structures~\cite{BESIII:2016bnd}. One (the $\psi(4230)$) has a slightly lower mass than the $Y(4260)$ and one (the $\psi(4360)$) has a higher mass.~The parameters of $\psi(4230)$ were measured in various processes, {\it i.e.} \mbox{$e^+e^-~\to~\omega\chi_{c0}$~\cite{BESIII:2014rja}}, $\pi^+\pi^-h_c$~\cite{BESIII:2016adj}, $\pi^+\pi^-\psi(3686)$~\cite{BESIII:2017tqk}, $D^{0}D^{*-}\pi^+$~\cite{BESIII:2018iea}, and so on.~However, decays of the exotic states into light hadrons have not been so far observed in the processes $e^+e^-\to K_{S}^{0}K^{\pm}\pi^{\mp}\pi^{0}(\eta)]$~\cite{BESIII:2018kyw}, $K_S^0K^{\pm}\pi^{\mp}$~\cite{BESIII:2018cco}, $p\bar{n}K_S^0K^- + c.c.$~\cite{BESIII:2018gvg}, $\phi\phi\phi$, $\phi\phi\omega$~\cite{BESIII:2017ftc}, $p\bar{p}\pi^0$~\cite{BESIII:2017qwj}, $p\bar{p}\eta(\omega)$~\cite{BESIII:2021vkt}, $2(p\bar{p})$~\cite{BESIII:2020svk}, $\omega\pi^{+}\pi^{-}$~\cite{BESIII:2023gqy}, and $\omega\pi^{0}(\eta)$~\cite{BESIII:2022zxr}.

While multiple theoretical approaches have been proposed to describe the exotic states, {\it e.g.} as tetraquarks, hadronic
molecules or hybrid charmonia, the inner structure of the exotic states, {\it e.g.} the $\psi(4230)$, is not fully understood.
Their compatibility with experimental data has recently been reviewed in detail in Ref.~\cite{Brambilla:2019esw}. Investigating the decay modes of resonances is an essential way to understand their properties and internal structures. Further exploration of light hadron final states is desirable to probe the nature of the charmonium-like states~\cite{Dubynskiy:2008mq, Close:2005iz}.

The process $e^+e^-\to~f_1(1285)\pi^+\pi^-$ has been studied only at the BaBar experiment, from threshold up to 4.550~GeV~\cite{BaBar:2007qju,BaBar:2022ahi}. In this work, we report the measurement of the process $e^+e^-~\to~f_1(1285)\pi^+\pi^-$, with $f_1(1285)~\rightarrow~\pi^+\pi^-\eta$ and $\eta\rightarrow\gamma\gamma$, based on data collected at the \mbox{BESIII} experiment at $45$ center-of-mass energies~($\sqrt{s}$) ranging from $3.808$ to $4.951$~GeV, with an integrated luminosity of $22.2~{\rm fb^{-1}}$~\cite{BESIII:2022ulv,BESIII:2022dxl,SongWm:2015lum}.

\section{\label{sec:level2}BESIII detector and Monte Carlo simulation}
The \mbox{BESIII} detector~\cite{BESIII:2009fln} records symmetric $e^+e^-$ collisions provided by the BEPCII storage ring~\cite{Yu:2016cof} in the center-of-mass energy range from $1.85$ to $4.95~{\rm GeV}$, with a peak luminosity of $1.1 \times 10^{33}\;\text{cm}^{-2}\text{s}^{-1}$ achieved at \mbox{$\sqrt{s} = 3.77\;\text{GeV}$}. \mbox{BESIII} has collected large data samples in this energy region~\cite{BESIII:2020nme,EventFilter}. The cylindrical core of the \mbox{BESIII} detector covers 93\% of the full solid angle and consists of a helium-based multilayer drift chamber~(MDC), a plastic scintillator time-of-flight system~(TOF), and a CsI(Tl) electromagnetic calorimeter~(\mbox{EMC}), all enclosed in a superconducting solenoidal magnet providing a $1.0~{\rm T}$ magnetic field. The solenoid is supported by an octagonal flux-return yoke with resistive plate counter muon identification modules interleaved with steel. 
The charged-particle momentum resolution at $1~{\rm GeV}/c$ is $0.5\%$, and the ${\rm d}E/{\rm d}x$ resolution is $6\%$ for electrons from Bhabha scattering. The \mbox{EMC} measures photon energies with a resolution of $2.5\%$ ($5\%$) at $1$~GeV in the barrel (end cap) region. The time resolution in the \mbox{TOF} barrel region is 68~ps, while that in the end cap region is 110~ps. The end cap \mbox{TOF} system was upgraded in 2015 using multigap resistive plate chamber technology, providing a time resolution of 60~ps~\cite{Li:2017tofupdate}. This improvement affects data at $30$ of the $45$ center-of-mass energy points.

Simulated data samples produced with a {\sc geant4}-based~\cite{GEANT4:2002zbu} Monte Carlo~(\mbox{MC}) package, which includes the geometric description of the \mbox{BESIII}~detector and the detector response, are used to determine detection efficiencies and to estimate backgrounds. The simulation models the beam energy spread and ISR in the $e^+e^-$ annihilations with the generator {\sc kkmc}~\cite{Jadach:1999vf}. Signal \mbox{MC} samples for the process $e^+e^-\rightarrow f_{1}(1285)\pi^+\pi^-$ with the subsequent decays $f_{1}(1285)\rightarrow\pi^+\pi^-\eta$, and $\eta\rightarrow\gamma\gamma$ are generated uniformly in phase space. The inclusive \mbox{MC} sample includes the production of open charm processes, the ISR production of vector charmonium(-like) states, and the continuum processes incorporated in {\sc kkmc}~\cite{Jadach:1999vf}. All particle decays are modeled with {\sc evtgen}~\cite{Lange:2001uf} using branching fractions either taken from the Particle Data Group (\mbox{PDG})~\cite{PDG}, when available, or otherwise estimated with {\sc lundcharm}~\cite{Chen:2000tv}. Final state radiation from charged final state particles is incorporated using the {\sc photos} package~\cite{Richter-Was:1992hxq}.

\section{\label{sec:level3}Event selection and background analysis}
The process $e^+e^-\to f_1(1285)\pi^+\pi^-$ is reconstructed with $f_1(1285)\to\pi^+\pi^-\eta$ and $\eta\to\gamma\gamma$. Therefore, the final state particles consist of $2(\pi^+\pi^-)$ and 2$\gamma$. Charged tracks detected in the MDC are required to be within a polar angle ($\theta$) range of $|\cos\theta|<0.93$, where $\theta$ is defined with respect to the $z$ axis, which is the symmetry axis of the MDC. The distance of closest approach to the interaction point (IP) must be less than 10\,cm along the $z$ axis, and less than 1\,cm in the transverse plane. Exactly four charged tracks with a net charge equal to zero is required. Particle identification~(\mbox{PID}) combines information from the energy loss in the \mbox{MDC} and the flight time in the \mbox{TOF} to calculate the probabilities for the $\pi$, $K$, and $p$ hypotheses. Each track is then assigned a particle type corresponding to the hypothesis with the highest probability. Events with four pion candidates are reserved.
 
Photon candidates are identified using isolated showers in the EMC. The deposited energy of each shower must be more than 25~MeV in the barrel region ($|\cos \theta|< 0.80$) and more than 50~MeV in the end cap region ($0.86 <|\cos \theta|< 0.92$). To exclude showers that originate from
charged tracks,
the angle subtended by the EMC shower and the position of the closest charged track at the EMC
must be greater than 10 degrees as measured from the IP. To suppress electronic noise and showers unrelated to the event, the difference between the EMC time and the event start time is required to be within 
[0, 700]\,ns. Events containing at least two photon candidates are then retained for further analysis. In order to further suppress backgrounds and improve the momentum resolution, a four-constraint kinematic fit based on four-momentum conservation is applied under the hypotheses of $\pi^+\pi^-\pi^+\pi^-\gamma\gamma$, with a requirement of $\chi^2 < 100$, where $\chi^{2}$ represents the goodness of the four-constraint kinematic fit. If there are more than two photon candidates in the event, the combination with the smallest $\chi^2$ is retained. 
 
The $\eta$ candidate is reconstructed by two photons with \mbox{$|M(\gamma\gamma)-M(\eta)|~<~25~{\rm MeV}$}, where $M(\gamma\gamma)$ is the invariant mass reconstructed from the two photons and $M(\eta)$ is the known mass of $\eta$~\cite{PDG}.

The invariant mass of $M(\pi^+\pi^-\eta)$ for all four combinations within each event is illustrated in \mbox{Fig}.~\ref{fitdata}, as an example, using the center-of-mass energy at $\sqrt{s}$ = 4.178 GeV that gives the highest statistics among the studied points. A clear $f_1(1285)$ signal is observed. Background contributions are studied through an analysis of the inclusive \mbox{MC} sample with an event type analysis tool TopoAna~\cite{Zhou:2020ksj}, indicating that the primary background originates from the continuum process \mbox{$e^+e^-\rightarrow \pi^+\pi^-\pi^+\pi^-\eta$}.

\section{\label{sec:level4}Cross section}
The Born cross section at each center-of-mass energy is determined by
\begin{equation}
	\sigma_{\rm B}~=~\frac{N_{\rm sig}}{\epsilon\cdot(1+\delta_\gamma)\cdot(1+\delta_v)\cdot\mathcal{B}\cdot\mathcal{L}},
	\label{func}
\end{equation}
where $1+\delta_{v} = \frac{1}{|1-\Pi|^2}$ accounts for vacuum polarization~(VP) corrections. The product branching fraction $\mathcal{B}=\mathcal{B}_{f_1(1285)}\mathcal{B}_{\eta}=13.8\%$ corresponds to the decay chain $f_1(1285)\rightarrow \pi^+\pi^-\eta$ and $\eta \rightarrow \gamma\gamma$. $N_{\rm sig}$ represents the signal yield extracted from the fit to the $M(\pi^+\pi^-\eta)$ distribution, $\mathcal{L}$ is the integrated luminosity~\cite{Goldhaber:1977qn,DASP:1978dns,Siegrist:1976br}, $1+\delta_{\gamma}$ is the ISR correction factor, and $\epsilon$ is the signal detection efficiency.
The dressed cross section is obtained by
\begin{equation}
	\sigma_{\rm d}~=~\sigma_{\rm B}\cdot(1+\delta_{v}),
	\label{funcdress}
\end{equation}
which incorporates the vacuum polarization correction.

\begin{figure}[tp]
	\centering
	\includegraphics[width=0.48\textwidth]{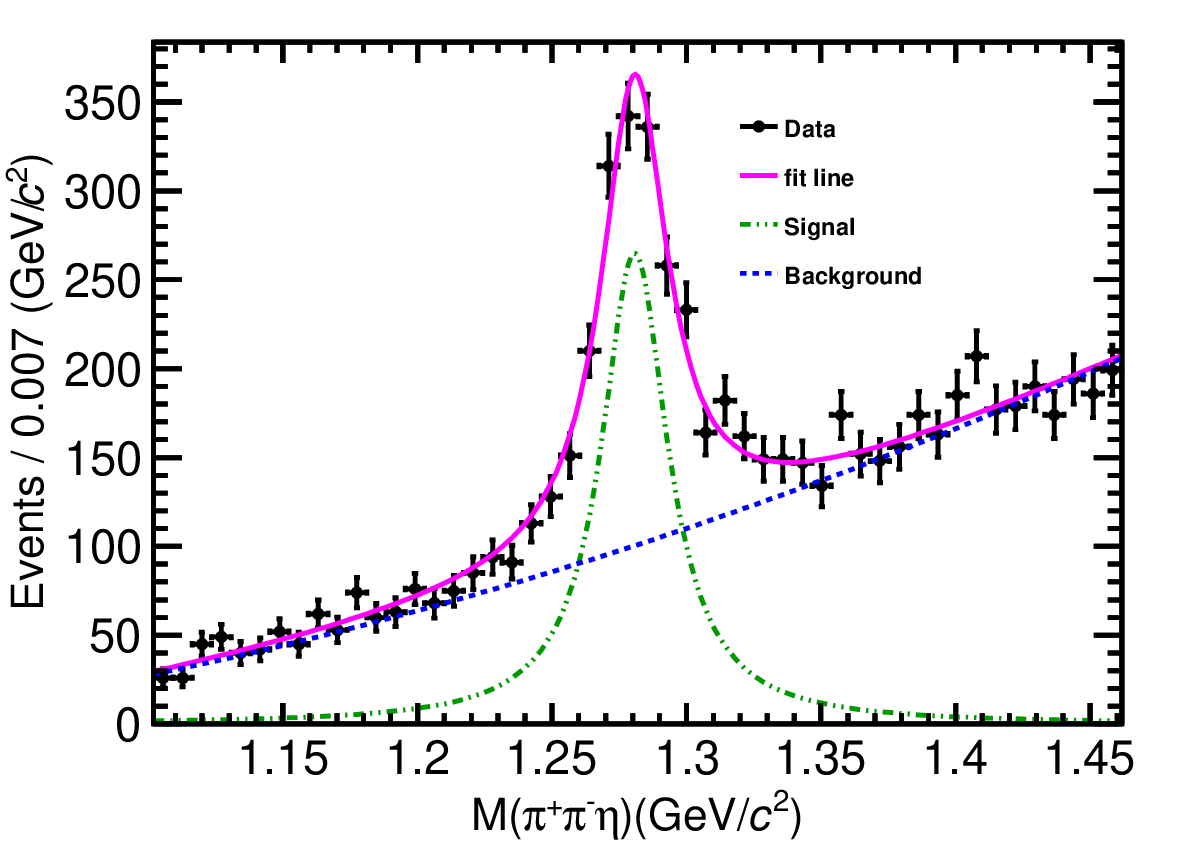}
	\setlength{\abovecaptionskip}{-0.34cm}
	\setlength{\belowcaptionskip}{-0.4cm}
	\caption{The $M(\pi^+\pi^-\eta)$ distribution for all four combinations within each event at $\sqrt{s}~=4.178~{\rm GeV}$. The black dots with error bars are data. The magenta solid line represents the fit result for signal extraction, the green dash-dotted line represents the $f_{1}(1285)$ shape, and the blue dashed line indicates the background shape.
	}
	\label{fitdata}
\end{figure}

The signal yield is obtained through an unbinned maximum likelihood fit to the $M(\pi^+\pi^-\eta)$ distribution. The signal shape is modeled by the \mbox{MC} shape convolved with a Gaussian function to compensate the detector resolution. The background shape is described by a second-order polynomial function. An example of the fit at $\sqrt{s}=4.178~{\rm GeV}$ is shown in \mbox{Fig}.~\ref{fitdata}. To prevent distortions from low statistics at other center-of-mass energies, the parameters of the convolved Gaussian function are fixed to the values obtained at $\sqrt{s}=4.178~{\rm GeV}$, with a mean $\mu=0.11$~$\rm MeV/c^{2}$ and width $\sigma=0.14~\rm MeV/c^{2}$. 

\begin{figure}[pb]
	\includegraphics[width=0.48\textwidth]{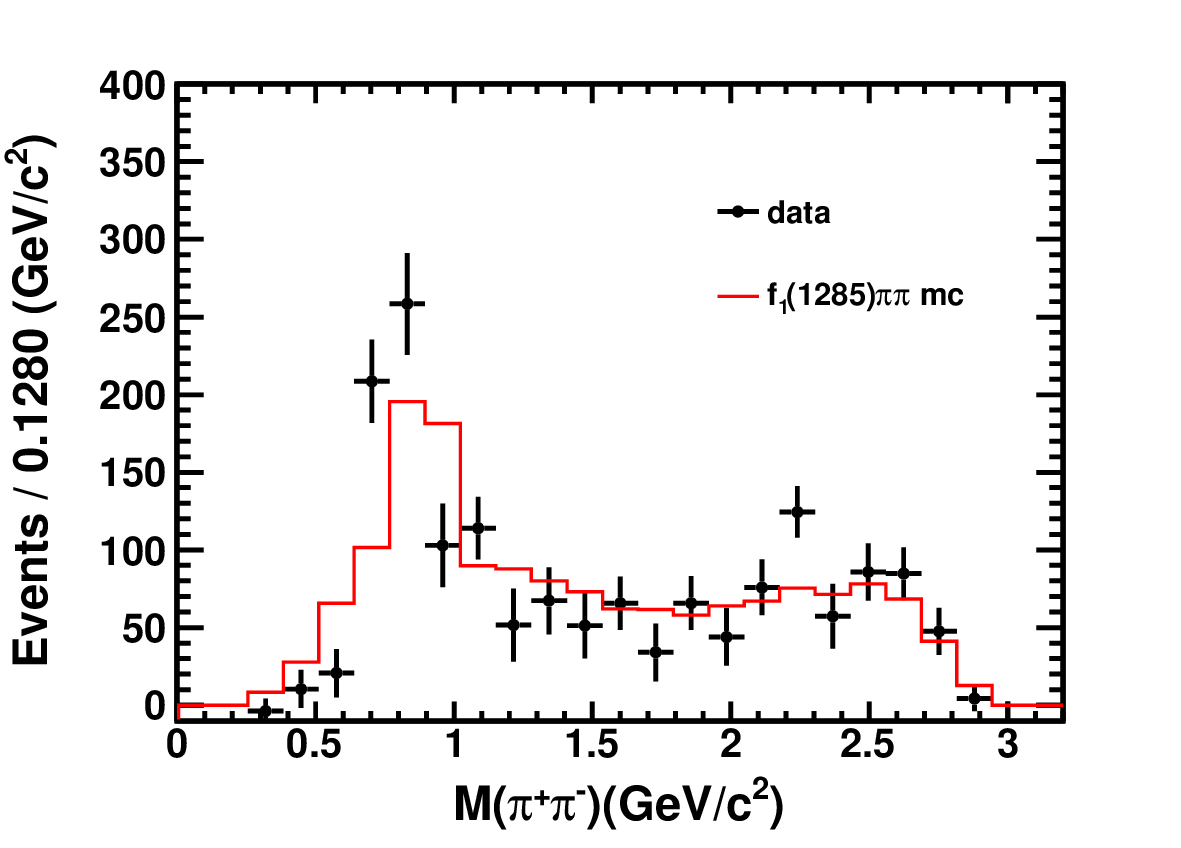}
	\vspace{-6pt}
	\setlength{\abovecaptionskip}{-0.14cm}
	\setlength{\belowcaptionskip}{-0.15cm}
	\caption{The distribution of $M(\pi^+\pi^-)$ from $e^+e^-$ collision at $\sqrt{s} = 4.178 \rm~GeV$. The black dots with error bars are data, the histogram denotes the weighted distribution from signal MC sample.}
	\label{figfitweight}
\end{figure}

Signal efficiency is obtained from the signal \mbox{MC} sample for $e^+e^-\to f_1(1285)\pi^+\pi^-$, with $f_1(1285)~\to~\pi^+\pi^-\eta$ and $\eta\to\gamma\gamma$.~To reduce discrepancies between data and signal MC, a weighting method is applied to the two-dimensional distributions of $M_{col}(f_1(1285)\pi^{+})$ versus $M_{col}(\pi^{+}\pi^{-})$ and $M_{f_1}(\eta\pi^{+})$ versus $M_{f_1}(\pi^{+} \pi^{-})$, where $\pi$s in $M_{col}$ and $M_{f_1}$ are from $e^{+}e^{-}$ collisions and $f_{1}(1285)$ decay, respectively. The $f_1(1285)$ candidates are selected as the $\pi^{+}\pi^{-}\eta$ combination with invariant mass closest to the value of $M_{f_{1}(1285)}$ among the four combinations in an event in both the data and signal \mbox{MC} samples. $M_{f_{1}(1285)}$ represents the known $f_{1}(1285)$ mass~\cite{PDG}. Figure~\ref{figfitweight} shows the comparison on $M(\pi^+\pi^-)$ from $e^+e^-$ collision between data and the weighted signal MC at $\sqrt{s} = 4.178 \rm~GeV$.

The correction factors $1+\delta_\gamma$ and $\epsilon$, which depend on the line shape, are obtained using an iterative method~\cite{BESIII:2021yam}. Both the Born and dressed cross sections follow the trend described by Eq.~(\ref{coneq}), as discussed in Section~\ref{sec:level6}. The iterative process begins with fits to the Born cross section line shape, starting at 3.7 GeV, below the lowest center-of-mass energy in this study. The $\epsilon$ and $(1+\delta_{\gamma})(1+\delta_{v})$ are calculated from the fit curve at each center-of-mass energy and used as input for the next iteration, repeating until the Born cross section converges.

The results of cross sections at each center-of-mass energy are presented in Table~\ref{tabxsvalue}, which includes both statistical and systematic uncertainties for the dressed cross sections ($\sigma_{\rm d}$). Detailed investigations of the systematic uncertainties are discussed in Section~\ref{sec:level5}.

\begin{center}
\begin{ruledtabular}
\setlength{\tabcolsep}{4.35mm}
\setlength{\LTcapwidth}{7in}
\begin{longtable*}{cccccccccc}
\caption{\footnotesize{Dressed cross sections of the process $e^+e^-\to f_1(1285)\pi^+\pi^-$. $\mathcal{L}$ is the integral luminosity of the data sample at each center-of-mass energy, $N_{\rm sig}$ is the number of signal events, $1+\delta_{\gamma}$ and $1+\delta_{v}$ are the converged radiation and vacuum polarization correction factor, $\epsilon$ is the signal efficiency with the weighting method, $\epsilon_{\rm phsp}$ for comparison is the signal efficiency directly acquired from signal MC sample, and $\sigma_{\rm d}$ is the dressed cross section. The first uncertainties are statistical and the second ones are systematic. Details of all systematic uncertainties are discussed in Section~\ref{sec:level5}. }}
\label{tabxsvalue}
			
			\\ \hline 
			$\sqrt{s}~(\rm {GeV})$& $\mathcal{L}~(\rm pb^{-1})$ & $N_{\rm sig}$  &$1+\delta_{\gamma}$ & $1+\delta_{v}$ &$\epsilon$ (\%) & $\epsilon_{\rm phsp}$ (\%) & $\sigma_{\rm d}$~($\rm pb$) \\						
			\endfirsthead

			\multicolumn{9}{r}{Continued}\\
			\hline

			$\sqrt{s}~(\rm {GeV})$ & $\mathcal{L} ~(\rm pb^{-1})$ & $N_{\rm sig}$ &$1+\delta_{\gamma}$ & $1+\delta_{v}$  &$\epsilon$ (\%) & $\epsilon_{\rm phsp}$ (\%)  & $\sigma_{\rm d}$~($\rm pb$) \\
			\hline
			\endhead 
			
		      \hline
                         \multicolumn{9}{r}{Continued on next page}\\
			\endfoot
			
			\endlastfoot

			\hline
3.808    &50.5     &56$\pm$12  &0.86 &1.06 &22.9  &24.2  &41.0$\pm$8.8$\pm$17.8\\
\hline
3.869    &219.2    &137$\pm$22  &0.89 &1.05  &20.9 &21.9 &24.3$\pm$3.9$\pm$10.5\\
\hline
3.896    &52.6     &42$\pm$11  &0.90 &1.05  &21.3   &22.5  &30.1$\pm7.9\pm$13.0\\
\hline
4.008    &482.0    &331$\pm$31  &0.94 &1.04 &19.9  &20.9 &26.6$\pm$2.5$\pm$11.5\\
\hline
4.085    &52.9     &40$\pm$10  &0.96 &1.05 &19.5   &20.4 &29.3$\pm$7.3$\pm$12.7\\
\hline
4.129    &401.5  &195$\pm$26   &0.98 &1.05 &19.9 &20.6 &18.2$\pm$2.4$\pm$7.9\\
\hline
4.158    &408.7  &179$\pm$25  &0.98 &1.05 &19.8 &20.5 &16.4$\pm$2.3$\pm$7.1\\
\hline
4.178    &3189.0  &1593$\pm$69 &0.99&1.05  &18.1 &18.6 &20.3$\pm$0.9$\pm$8.8\\
\hline
4.189
&526.7  &235$\pm$27   &0.99&1.06 &18.4 &18.8 &17.8$\pm$2.0$\pm$7.7\\
\hline
4.199
&526.0  &268$\pm$27  &0.99&1.06 &18.6 &19.0 &20.1$\pm$2.0$\pm$8.7\\
\hline
4.209
&517.1 &227$\pm$27  &0.99&1.06 &18.2 &18.6 &17.6$\pm$2.1$\pm$7.6\\
\hline
4.219
&514.6 &280$\pm$28 &1.00&1.06 &18.3 &18.6 &21.7$\pm$2.2$\pm$9.4\\
\hline
4.226
&1100.9  &494$\pm$40 &1.00&1.06  &18.5 &19.3 &17.6$\pm$1.4$\pm$7.6\\
\hline
4.236
&530.3 &248$\pm$27  &1.00&1.06 &18.6 &18.9 &18.3$\pm$2.0$\pm$7.9\\
\hline
4.242
&55.9  &36$\pm$10  &1.00&1.06 &18.5 &19.2 &25.3$\pm$7.0$\pm$10.9\\
\hline
4.244
&538.1  &212$\pm$26 &1.00&1.06  &18.5 &18.8 &15.4$\pm$1.9$\pm$6.7\\
\hline
4.258
&828.4 &359$\pm$33  &1.01&1.05 &18.2  &18.9 &17.2$\pm$1.6$\pm$7.4\\
\hline
4.267
&531.1  &259$\pm$27  &1.01&1.05 &18.5 &18.8 &19.0$\pm$2.0$\pm$8.2\\
\hline
4.278
&175.7 &67$\pm$15 &1.01&1.05  &18.3 &18.4 &15.0$\pm$3.4$\pm$6.5\\
\hline
4.288
&502.4  &221$\pm$25 &1.01&1.05  &19.2 &19.6&16.5$\pm$1.9$\pm$7.1\\
\hline
4.308
&45.1 &23$\pm$8  &1.02&1.05 &17.9 &18.5 &20.4$\pm$7.1$\pm$8.8\\
\hline
4.313
&501.2 &244$\pm$26 &1.02&1.05 &19.1 &19.5 &18.2$\pm$1.9$\pm$7.9\\
\hline
4.338
&505.0  &189$\pm$25 &1.02&1.05  &19.4 &19.7 &13.7$\pm$1.8$\pm$5.9\\
\hline
4.358
&544.0 &204$\pm$25 &1.03&1.05  &17.8 &18.4 &14.9$\pm$1.8$\pm$6.4\\
\hline
4.378
&522.7  &201$\pm$25 &1.03&1.05  &19.1 &19.5 &14.2$\pm$1.8$\pm$6.1\\
\hline
4.387
&55.6 &28$\pm$9 &1.03&1.05  &17.5 &18.2 &20.2$\pm$6.5$\pm$8.8\\
\hline
4.397
&507.8 &198$\pm$24 &1.03&1.05  &19.1  &19.3 &14.3$\pm$1.7$\pm$6.2\\
\hline
4.416
&1043.9  &384$\pm$34  &1.04&1.05 &18.8 &19.5 &13.7$\pm$1.2$\pm$5.9\\
\hline
4.437
&569.9 &173$\pm$25 &1.04&1.05 &18.7 &19.4 &11.3$\pm$1.6$\pm$4.9\\
\hline
4.467
&111.1 &51$\pm$12 &1.05&1.05 &18.6 &19.3 &17.1$\pm$4.0$\pm$7.4\\
\hline
4.527
&112.1 &39$\pm$9 &1.06&1.05 &18.5 &19.1 &12.9$\pm$3.0$\pm$5.6\\
\hline
4.575
&48.9 &$3^{+4}_{-3}$ &1.07&1.05 &18.6 &19.0 &2.2$^{+3.0}_{-2.2}\pm$1.0\\
\hline
4.600
&586.9 &170$\pm$22 &1.07&1.05 &18.4 &18.8 &10.6$\pm$1.4$\pm$4.6\\
\hline
4.612
&103.7 &17$\pm$8 &1.08&1.05 &17.7 &18.1 &6.2$\pm$2.9$\pm$2.7\\
\hline
4.628
&521.5 &183$\pm$22  &1.08&1.05 &17.4 &18.0 &13.6$\pm$1.6$\pm$5.9\\
\hline
4.641
&551.7  &126$\pm$20  &1.08&1.05 &17.7 &18.0 &8.7$\pm$1.4$\pm$3.8\\
\hline
4.661
&529.4 &141$\pm$20 &1.09&1.05 &17.4 &17.9 &10.2$\pm$1.5$\pm$4.4\\
\hline
4.682
&1667.4  &400$\pm$35 &1.09&1.05  &17.4  &17.8 &9.2$\pm$0.8$\pm$4.0\\
\hline
4.699
&535.5 &113$\pm$18 &1.09&1.05  &17.3 &17.7 &8.1$\pm$1.3$\pm$3.5\\
\hline
4.740
&163.9 &37$\pm$10  &1.10&1.05 &17.4 &17.8 &8.6$\pm$2.3$\pm$3.7\\
\hline
4.750
&366.6 &89$\pm$16  &1.10&1.05 &17.4 &17.8 &9.2$\pm$1.7$\pm$4.0\\
\hline
4.781
&511.5 &100$\pm$17  &1.11&1.06&17.3 &17.6 &7.4$\pm$1.3$\pm$3.2\\
\hline
4.843
&525.2 &105$\pm$18  &1.12&1.06 &17.1 &17.6 &7.6$\pm$1.3$\pm$3.3\\
\hline
4.918
&207.8 &36$\pm$9  &1.13&1.06 &16.7 &17.3 &6.7$\pm$1.7$\pm$2.9\\
\hline
4.951
&159.3 &6$^{+8}_{-7}$ &1.14&1.06 &16.6 &17.0 &1.4$^{+1.9}_{-1.7}\pm$0.6\\
\hline

		\end{longtable*}
		\end{ruledtabular}
\end{center}

\section{\label{sec:level5}Systematic uncertainties}
The systematic uncertainties are estimated at each center-of-mass energy. Multiple sources of systematic uncertainties are taken into account in the measurement of the Born cross section for the $e^+e^-~\rightarrow~f_{1}(1285)\pi^+\pi^-$ process. These include the uncertainties from integrated luminosity measurement, tracking and \mbox{PID}, photon reconstruction, quoted branching fractions, kinematic fit, signal yield estimation, ISR and VP correction factors, as well as uncertainties in signal efficiencies. 

The integrated luminosity is determined using large angle Bhabha events, with an uncertainty of $1\%$~\cite{BESIII:2022ulv,BESIII:2022dxl,SongWm:2015lum}. The systematic uncertainty of the tracking and \mbox{PID} efficiencies are both measured as $1\%$ per track by using control samples $J/\psi \rightarrow K^{0}_{S}K^{\pm}\pi^{\mp}$, $J/\psi \rightarrow p\bar{p}\pi^{+}\pi^{-}$, and $J/\psi \rightarrow \pi^+\pi^-\pi^0$~\cite{BESIII:2017qwj,BESIII:2013qmu,BESIII:2010ank,Yuan:2015wga,BESIII:2012urf,BESIII:2011ysp}. A systematic uncertainty of $0.2\%$ is assigned to the single photon reconstruction efficiency using a $J/\psi\rightarrow\gamma\mu^+\mu^-$ control sample. Additionally, the systematic uncertainties of the branching fractions are quoted from the \mbox{PDG}~\cite{PDG}, with values of $42.9\%$ and $0.5\%$ for $f_{1}(1285) \rightarrow \eta\pi^+\pi^-$ and $\eta \rightarrow \gamma\gamma$, respectively.

The systematic uncertainty of the kinematic fit is determined by taking into account the track helix correction parameters in the signal \mbox{MC} sample. The value of $\pi$ type track correction parameter is obtained from the control sample $e^+e^- \rightarrow \pi^{+}\pi^{-}J/\psi$. The difference in efficiency from the nominal result is considered as the systematic uncertainty caused by the kinematic fit~\cite{BESIII:2012mpj}.

Three sources of the systematic uncertainties are taken into account for signal yield estimation: signal shape, background shape, and fit range. To mitigate the impact of statistical fluctuation, samples from all center-of-mass energies are combined for the systematic study. The discrepancy in the signal yields obtained by using the Breit-Wigner function convolved with a Gaussian instead of the \mbox{MC} shape convolved with a Gaussian function is attributed to the systematic uncertainty arising from the signal shape description. Similarly, the difference in the signal yields when using a third-order polynomial to describe the background shape, as opposed to the nominal method, is taken as the systematic uncertainty stemming from the background shape description. The systematic uncertainty associated with the fit range is determined by comparing the results obtained when enlarging or shrinking the fit range by $10~{\rm {MeV}}$ on both sides, with the largest difference compared to the nominal signal yield being considered as the systematic uncertainty.

The systematic uncertainty associated with ISR is assessed from two perspectives: the iterative method and the choice of the start point for the line shape fit. By randomly varying the parameters of the converged line shape for a sufficient number of times (in this analysis, 1000 times are used), within 1$\sigma$ of each parameter value, a distribution of the new $\epsilon(1+\delta_{\gamma})(1+\delta_{v})$ with 1000 entries is obtained. This distribution is then fitted by a Gaussian function. The ratio of the width to the mean of the fitted Gaussian distribution is considered as the systematic uncertainty related to the iterative method.

To determine the systematic uncertainty related to the choice of the starting point in the line shape fit, the start point is set to 1.8 GeV, close to the production threshold of the $e^+e^-\rightarrow f_{1}(1285)\pi^+\pi^-$ process. For energies below 3.7~$\rm GeV$, the results from the BaBar experiment are utilized. Given that {\sc kkmc}~\cite{Jadach:1999vf} is originally designed for simulating $c\bar{c}$ processes and may not accurately represent processes below the $c\bar{c}$ threshold, the signal \mbox{MC} samples are generated using the {\sc conexc} generator~\cite{Ping:2013jka} to address the deficiency. The differences between the two generators are found to be less than $0.2\%$, which is considered negligible. The systematic uncertainty from the line shape fit range is deteremined by comparing the $\epsilon(1+\delta_{\gamma})(1+\delta_{v})$ values obtained from the complete line shape fit with the nominal values.

The systematic uncertainty associated with the signal efficiency, derived from the weighting method, is assessed by varying the weights of the two-dimensional histograms within the bin error. This process is repeated 1000 times to generate a distribution of efficiencies. A Gaussian function is fitted to this distribution, and the systematic uncertainty is determined by the ratio of the width to the mean of the fitted Gaussian distribution.

The total systematic uncertainty at each center-of-mass energy is calculated by summing the individual contributions in quadrature. These systematic uncertainties are then propagated to the dressed cross sections using the relation given by Eq.~(\ref{funcdress}).

\section{\label{sec:level6} The fit of $e^+e^-\rightarrow f_{1}(1285)\pi^+\pi^-$ cross section}

\begin{figure}[bp]
	\centering
	\includegraphics[width=0.48\textwidth]{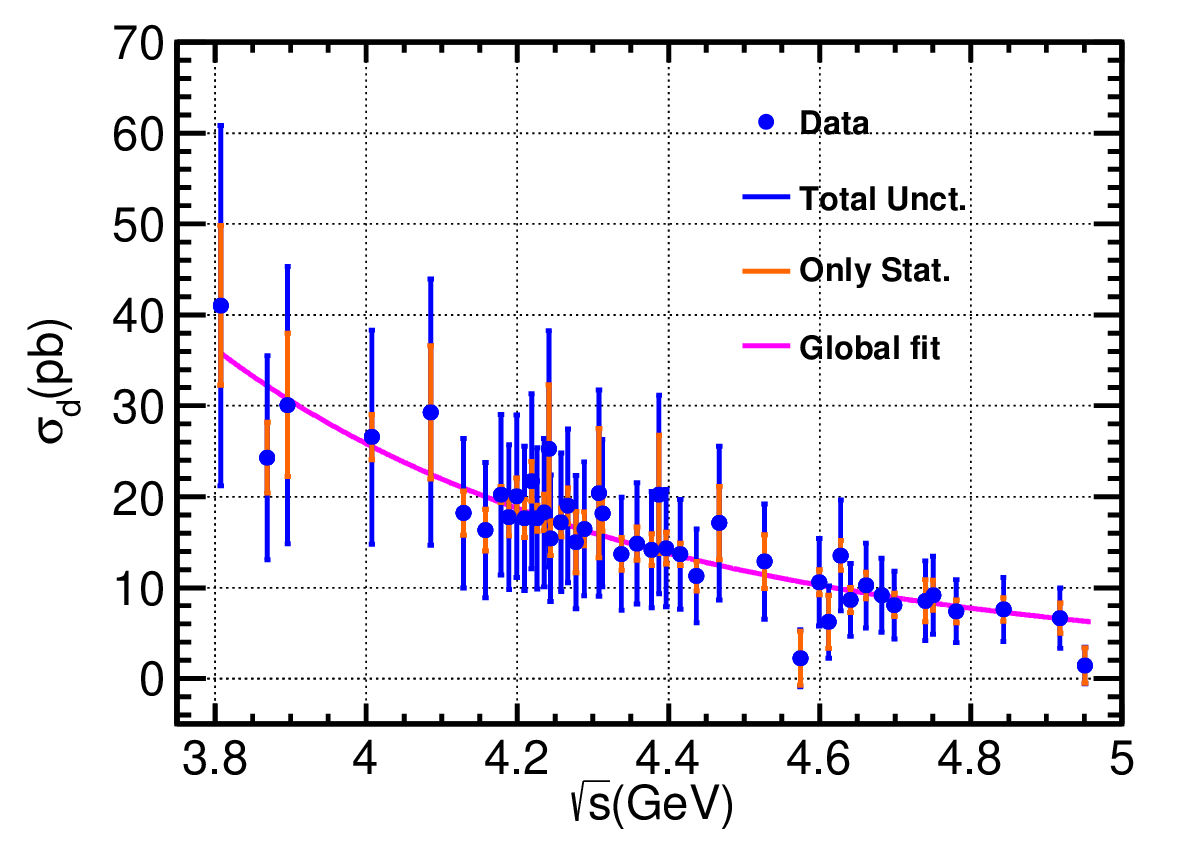}
		\setlength{\abovecaptionskip}{-0.70cm}
		\setlength{\belowcaptionskip}{-0.35cm}
	\vspace{8pt}
	\caption{ The line shape of the cross section of the process $e^+e^-\rightarrow f_{1}(1285)\pi^+\pi^-$ described by a continuum process. The blue points are the measured central values of the dressed cross section, the blue bars represent the total statistical and systematic uncertainties, and the orange bars represent the statistical uncertainty.
		\label{ccon}
	}
\end{figure}

The dressed cross section of the process $e^+e^-\rightarrow f_{1}(1285)\pi^+\pi^-$ is shown in \mbox{Fig}.~\ref{ccon}, with statistical and systematic uncertainties summed in quadrature. No obvious resonance is observed in the energy region from $\sqrt{s}=3.808$ to $4.951~{\rm GeV}$. The line shape can be described by \cite{BESIII:2017qwj,BESIII:2023dhc}

\begin{equation}
	\sigma_{\rm d}~=~(\frac{c}{\sqrt{s}})^{\lambda},
	\label{coneq}
\end{equation}

\noindent where $c$ and $\lambda$ are free parameters to be determined in the fit. 

A minimization $\chi^2$ in an unbiased definition~\cite{Mo:2003cna,Mo:2003jwa,Mo:2006bea,Mo:2007aea} is 

\begin{equation}
	\chi^2~=~\sum_{i}^{n}(\frac{\sigma_{i}-f\sigma^{fit}_{i}}{\delta_{i}})^2+(\frac{1-f}{\delta_{f}})^2,
	\label{chisq_func}
\end{equation}

\noindent where $\sigma_{i}$ and $\sigma^{fit}_{i}$ are the measured and fitted dressed cross sections at the $i$th energy point, respectively. $\delta_{i}$ is the uncorrelated uncertainty, $f$ is the scale factor as a free parameter in the fit, and $\delta_{f}$ is the relative systematic uncertainty of $f$, determined by the total correlated uncertainties. Systematic uncertainties are treated as correlated, while the statistical uncertainties are uncorrelated. The correlated systematic uncertainties are estimated at each center-of-mass energy, resulting in slight differences at different $\sqrt{s}$. The value at $\sqrt{s}=4.178~\rm GeV$ is taken as the $\delta_{f}$ in the fit. The fit result is shown in \mbox{Fig}.~\ref{ccon}. The goodness of the fit is indicated by $\chi^2/ndf = 47.3/42 = 1.1$. The parameter $\lambda$ denotes the trend of cross section with respect to center-of-mass energy. The values of $\lambda = 6.61\pm 0.37 \pm 0.27$ and $c = 6.54 \pm 0.49 \pm 0.55$ are determined from the fit, where the first uncertainties are statistical and the second are systematic. Systematic uncertainties for $\lambda$ and $c$ are derived from the cross-section measurement and the fit range of the line shape. 

To explore the possible resonance contribution, a Breit-Wigner function is added coherently to Eq.~(\ref{coneq}),

\begin{equation}
	\sigma_{\rm d}~=~ [\sqrt{({\frac{c}{\sqrt{s}}})^{\lambda}} + \frac{\sqrt{12\pi C\varGamma_{ee}\mathcal{B}_{f_{1}(1285)\pi^+\pi^-}\varGamma}}{s - M^2 + iM\varGamma}e^{i\phi}]^2,
	\label{func2}
\end{equation}
\noindent where $\varGamma_{ee}$, $\mathcal{B}_{f_{1}(1285)\pi^+\pi^-}$, $M$, $\varGamma$, $s$, $\phi$ and $C$ are the partial width of the resonance decaying to $e^+e^-$, the branching fraction of the resonance decaying to $f_{1}(1285)\pi^+\pi^-$, the mass of the resonance, the width of the resonance, the square of the center-of-mass energy, the relative phase of the resonance compared to the non-resonant process, and a conversion constant equal to $3.893\times10^{5}~\rm nb~\rm GeV^2$, respectively. In the fit, $M$ and $\varGamma$ are fixed to the known masses and widths of the resonances $\psi(4040), \psi(4160), \psi(4230), \psi(4360), \psi(4415)$, or $\psi(4660)$~\cite{PDG}, while other parameters are left free to vary. The fit is performed individually for each resonance, and the significance of all possible resonances is found to be less than $2\sigma$ in each case.

\section{\label{sec:level7}summary}

The cross sections of \mbox{$e^{+}e^{-}\rightarrow~f_{1}(1285)\pi^{+}\pi^{-}$} at center-of-mass energies from 3.808 to 4.951 GeV are reported. The precision is improved by a factor of two compared to the previous results~\cite{BaBar:2007qju,BaBar:2022ahi}, and the first cross section measurement is provided in a range extending from 4.550 GeV up to 4.951 GeV. From the analysis of the line shape as shown in \mbox{Fig}.~\ref{ccon}, no significant resonance is observed. The energy-dependent cross section of $e^+e^-\rightarrow f_{1}(1285)\pi^+\pi^-$ can be well described by a power law function, with a goodness-of-fit $\chi^{2}/ndf = 47.3/42 = 1.1$. Individual examinations of potential resonance contributions from the charmonium states $\psi(4040)$, $\psi(4160)$, $\psi(4230)$, $\psi(4360)$, $\psi(4415)$, and $\psi(4660)$ indicate that their significances are each below the 2$\sigma$ threshold.

\section*{Acknowledgement}
We are grateful to Dr. Zhilong Han from University of Jinan for his discussions. The BESIII Collaboration thanks the staff of BEPCII and the IHEP computing center for their strong support. This work is supported in part by National Key R\&D Program of China under Contracts Nos. 2020YFA0406300, 2020YFA0406400, 2023YFA1606000; National Natural Science Foundation of China (NSFC) under Contracts Nos. 11635010, 11735014, 11935015, 11935016, 11935018, 12025502, 12035009, 12035013, 12061131003, 12192260, 12192261, 12192262, 12192263, 12192264, 12192265, 12221005, 12225509, 12235017, 12361141819; the Chinese Academy of Sciences (CAS) Large-Scale Scientific Facility Program; the CAS Center for Excellence in Particle Physics (CCEPP); Joint Large-Scale Scientific Facility Funds of the NSFC and CAS under Contract No. U1832207; 100 Talents Program of CAS; The Institute of Nuclear and Particle Physics (INPAC) and Shanghai Key Laboratory for Particle Physics and Cosmology; German Research Foundation DFG under Contracts Nos. 455635585, FOR5327, GRK 2149; Istituto Nazionale di Fisica Nucleare, Italy; Ministry of Development of Turkey under Contract No. DPT2006K-120470; National Research Foundation of Korea under Contract No. NRF-2022R1A2C1092335; National Science and Technology fund of Mongolia; National Science Research and Innovation Fund (NSRF) via the Program Management Unit for Human Resources \& Institutional Development, Research and Innovation of Thailand under Contract No. B16F640076; Polish National Science Centre under Contract No. 2019/35/O/ST2/02907; The Swedish Research Council; U. S. Department of Energy under Contract No. DE-FG02-05ER41374. This paper is also supported by the NSFC under Contracts Nos. 11335008, 11605074, 12105276, and Joint Large-Scale Scientific Facility Funds of the NSFC and CAS under Contract Nos.  U1732263, U2032115. 



\end{sloppypar}
\end{document}